\documentclass[prd,a4,aps,showpacs,epsf,floats,amsfonts,amssymb,amsmath,nofootinbib]{revtex4}

\newcommand{\be}{\begin{equation}}\newcommand{\ee}{\end{equation}}
\newcommand{\bea}{\begin{eqnarray}}\newcommand{\eea}{\end{eqnarray}}
\newcommand{\brr}{\begin{array}}\newcommand{\err}{\end{array}}
\newcommand{\bit}{\begin{itemize}}\newcommand{\eit}{\end{itemize}}
\newcommand{\ben}{\begin{enumerate}}\newcommand{\een}{\end{enumerate}}

\def\lab{\label}
\def\lan{\langle}
\def\lf{\left}

\def\non{\nonumber}
\def\pa{\partial}\def\ran{\rangle}
\def\rar{\rightarrow}
\def\ri{\right}

\def\al{\alpha}\def\bt{\beta}

\def\te{\theta}

\def\om{\omega}

\def\1{{_{1}}}\def\2{{_{2}}}
\def\Omm{\Omega_{-}}
\def\Omp{\Omega_{+}}
\def\Ommk{\Omega_{-}^{k}}
\def\Ompk{\Omega_{+}^{k}}
\def\nof{:\;\!\!\;\!\!:}

\begin{document}

\title{Non-cyclic phases for neutrino oscillations in quantum field theory}

\author{Massimo Blasone${}^{\flat}$\footnote{Corresponding author. Email: mblasone@unisa.it}, Antonio Capolupo${}^{\flat}$,
Enrico Celeghini${}^{\sharp}$ and Giuseppe Vitiello${}^{\flat}$
}

\vspace{2mm}

\address{${}^{\flat}$ Dipartimento di Matematica e Informatica and INFN,
 Universit\`a di Salerno,  Fisciano (SA) - 84084 Italy,
\\ ${}^{\sharp}$ Dipartimento di  Fisica and INFN, Universit\`a di Firenze,
I-50019 Sesto Fiorentino (FI), Italy}

\begin{abstract}

We show the presence of non-cyclic phases for oscillating neutrinos
in the context of quantum field theory. Such phases carry information about
the non-perturbative
vacuum structure associated with the field mixing. By subtracting
the condensate contribution of the flavor vacuum, the previously
studied quantum mechanics geometric phase is recovered.

\end{abstract}

\pacs{14.60.Pq; 03.65.Vf; 11.10.-z}

%\keywords{Neutrino Oscillations, Geometric Phases, Quantum Field Theory}

\maketitle

\section{Introduction}

Much attention and study has been devoted in recent years to the
phenomenon of neutrino mixing and oscillations since it offers the
possibility to investigate new physics beyond the Standard Model of
elementary particle physics, also involving hot issues in
astro-particle physics and cosmology
\cite{Bilenky:1978nj,Bilenky,Mohapatra:1991ng}. As a matter of fact,
great experimental and theoretical achievements have been obtained
and new horizons have been opened to be explored in future research.
The mixing phenomenon also offers some  features such as its
connection with vacuum structure \cite{BV95,BHV98,BPT02} and dark energy
\cite{Capolupo:2006et},  which certainly deserve further study and
attention due to their physical relevance. Among such specific
features, the one of the geometric phase
\cite{Berry:1984jv,Aharonov:1987gg,Anandan:1990fq} characterizing
mixed neutrino evolution has been pointed out in
Ref.\cite{Blasone:1999tq} in the quantum mechanical (QM) framework
(Pontecorvo's formalism) of neutrino mixing
\cite{Bilenky:1978nj,Bilenky,Mohapatra:1991ng}.

In general, geometrical phases appear in many physical systems as an
observable characterization of the system evolution. The
phenomenological interest in the geometric phase in neutrino
evolution arises since it is found \cite{Blasone:1999tq} to be
function only of the mixing angle which thus can be measured (at
least in principle) independently from dynamical parameters such as
masses and energies.

Other aspects of geometric phases associated to neutrino
oscillations have been studied in Refs.\cite{gaetano,He:2004zc}.
%The result of Ref.\cite{Blasone:1999tq} has been obtained in the
%quantum mechanical (QM) framework (Pontecorvo's formalism) of
%neutrino mixing \cite{Bilenky:1978nj,Bilenky,Mohapatra:1991ng}.
The generalization \cite{Pancharatnam,Samuel,Polavieja} of geometric
phase to non-cyclic evolution, such as the case of three and four
flavor mixing, has been also recently analyzed by using the QM
formalism \cite{Wang:2000ep,Law:2007fb}. Such a formalism is known
to be an useful approximation of the quantum field theory (QFT)
formalism which provides the correct theoretical setting for the
study of particle mixing and oscillations
\cite{BV95,BHV98,BPT02,Fujii:1999xa,JM01,yBCV02,Blasone:2005ae,Blaspalm,BCRV01}.

Aim of the present paper is to study the Aharonov-Anandan geometric
invariant in neutrino evolution in such a QFT formalism. We show
that the QFT condensate leads to a non-cyclic time evolution of the
flavor states and we compute the non-cyclic phases for oscillating
neutrinos. The QM geometric phase is recovered by subtracting from
the Hamiltonian the contributions from the vacuum condensate. Some
light on both, the condensate structure of the vacuum of QFT
neutrino mixing, and its quantum mechanical approximation is thus
shed. Here we consider the case of two-flavor Dirac neutrino fields,
although the conclusions we reach can be  extended to the case of
three flavors \cite{yBCV02} and Majorana neutrinos \cite{Blaspalm} and to the case of mixed bosons
\cite{BCRV01}.

The paper is organized as follows. In Section II we summarize the
results on the geometric invariant obtained for oscillating
neutrinos in the context of QM. We show that the geometric phase
represents the distance along the evolution of the neutrino in the
projective Hilbert space, as measured by the Fubini-Study metric. In
Section III we study the geometric invariant and the non-cyclic
phases for neutrino oscillations in the context of QFT and Section
IV is devoted to discussions and conclusions. A brief summary of the
vacuum structure for  Dirac neutrino mixing is presented in Appendix
A. In Appendix B are reported useful formulas.

\section{Geometry of neutrino oscillations in Quantum Mechanics}

We want to study the Aharonov-Anandan geometric invariant
\cite{Anandan:1990fq}
\bea\label{sQM1} s = 2\,\int_{0}^{t} \Delta E \, dt' \, \eea
in the case of  neutrino mixing. In this Section we summarize the
results obtained in ref. \cite{Blasone:1999tq} in the QM formalism
for two Dirac neutrinos (the case of three neutrinos is discussed in
\cite{Blasone:1999tq,Wang:2000ep,Law:2007fb} and will be commented
upon in the following). We also study the invariant $s$ in terms of
Fubini-Study metrics. The mixing transformations are:
\bea\label{mixQM} |\nu_{e} \rangle & = & \cos \theta
\,|\nu_{1}\rangle  +   \sin \theta\,|\nu _{2}\rangle\,,
\\[2mm]
|\nu_{\mu} \rangle & = & - \sin \theta\, |\nu_{1}\rangle  +  \cos
\theta\,|\nu _{2}\rangle \,. \eea
We focus our attention on the electron neutrino. Same discussion
applies to the muon neutrino. For simplicity of notation, we omit
the momentum suffix ${\bf k}$, the helicity label $r$ and use $\hbar
= 1$ whenever no ambiguity arises. In the present case  $\Delta E$
in Eq. (\ref{sQM1}) is given by
\bea\label{:H:e-mu} \Delta E \equiv \Delta E_{e,\mu}\, =
\,\langle\nu_{e}(t)|\, H\,|\nu_{\mu}(t)\rangle &=&
\,\langle\nu_{\mu}(t)|\,
H\,|\nu_{e}(t)\rangle %= \Omm \sin\theta\,\cos\theta\,,
\eea
where $|\nu_{e}(t)\rangle$ and $|\nu_{\mu}(t)\rangle$ are the
electron and the muon neutrino states at time $t$.  $|\nu_{e}(t)
\rangle $ is given by
\bea  |\nu_{e}(t)\ran  \equiv ~ e^{-i  H  t} |\nu_{e}(0)\ran
%\\ \non \\ \non
= e^{-i \om_{\1} t} \lf(\cos\te\;|\nu_{\1}\ran \;+\; e^{-i \Omm t}\;
\sin\te\; |\nu_{\2}\ran \; \ri)~, \eea
where $\Omm \equiv \omega_{2} - \omega_{1}$,  $H |\nu_i\ran =
\omega_i |\nu_i\ran$~ and $\omega_i$ are the energies associated
with the mass eingestates $|\nu_i\ran$, with $i=1,2$. Since at a
time $T= 2\pi / \Omm$, the state is the same as  the original one,
apart from a phase factor:
\bea |\nu_{e}(T)\ran = e^{i \phi} |\nu_{e}(0)\ran ~,
\;\;\;\;\;\;\;\;\;\;\;\;\;\; \phi= - \,2\pi \om_{\1}/ \Omm ~, \eea
one obtains \cite{Blasone:1999tq} for $t = nT$
\bea\label{sQM} s = 2\,\int_{0}^{nT} \Delta E_{e,\mu} \, dt = 2 n
\pi \sin 2 \theta\,. \eea
which is a function of the mixing angle only. We remark that the
same result is obtained by observing that the phase $\phi$ contains
a dynamical part and a geometric part $\beta_{e}$
\cite{Anandan:1990fq}
\bea \lab{Berry} { \bt_e} &=& \phi + \int_{0}^{T}\;\lan
\nu_{e}(t)|\;i\pa_t\;|\nu_{e}(t)\ran = { 2 \, \pi\; \sin^2\te}~,
\eea
For muon neutrinos we get $ { \bt_\mu} \,=\,  { 2 \, \pi\;
\cos^2\te}~. $
We have $\bt_e \, + \,\bt_\mu \, = \, 2\pi$. By considering the time
interval $(0,nT)$, one thus sees that  the geometric phase counts
oscillations, i.e. after $n$ oscillations the phase of the electron
neutrino state is $\,2 \,\pi \, n \;\sin^2\te$.

Let us now analyze the invariant $s$ in terms of the distance
between states in the Hilbert space. The evolution of the Pontecorvo
states $|\nu_{\sigma}(t) \rangle $, is governed by the Schr\"odinger
equation
\bea i\hbar \frac{d}{dt
}|\nu_{\sigma}(t) \rangle =\, H |\nu_{\sigma}(t) \rangle \,,
\qquad\qquad\, \sigma = e, \mu\,. \eea
 Expanding the state
$|\nu_{\sigma}(t+ dt) \rangle$ up to the second order in $dt$ and
considering that $\frac{d}{dt}H ~=~0$, we have %%
%\bea |\nu_{\sigma}(t+ dt) \rangle =
%|\nu_{\sigma}(t) \rangle - \frac{i dt}{\hbar}:H: |\nu_{\sigma}(t)
%\rangle - \frac{dt^{2}}{2 \hbar} \lf(i (\frac{d}{dt}:H:)
%|\nu_{\sigma}(t) \rangle +\frac{1}{\hbar} (:H:)^{2}|\nu_{\sigma}(t)
%\rangle \ri) +O(dt^{3})\,. \eea
%%
%
\bea \,\langle \nu_{\sigma}(t)
|\nu_{\sigma}(t+ dt) \rangle = 1 - \frac{i dt}{\hbar}
 \,\langle \nu_{\sigma}(t) | H |\nu_{\sigma}(t) \rangle  - \frac{dt^{2}}{2 \hbar^{2}}
\,\langle \nu_{\sigma}(t) | H^{2}|\nu_{\sigma}(t) \rangle +
O(dt^{3})\,, \eea
and
\bea\label{modulusQM}
|\langle \nu_{\sigma}(t) |\nu_{\sigma}(t+ dt) \rangle|^{2} =1-\frac{dt^{2}}{\hbar^{2}}
\Delta E_{\sigma,\sigma} ^{2}\, + \, O(dt^{3})\,,
\eea
where
\bea\label{deltaQM}
 \Delta E_{\sigma,\sigma}^{2}\, \equiv  \,\langle\nu_{\sigma}(t)|\,
 H ^{2}\,|\nu_{\sigma}(t)\rangle
- \langle\nu_{\sigma}(t)|\,
 H \,|\nu_{\sigma}(t)\rangle^{2} = (\Omm )^{2} \sin^{2}\theta\,\cos^{2}\theta\,,
 \qquad \sigma = e, \mu\,.
\eea
Then, we obtain
\bea \lf|\langle \nu_{\sigma}(t)
|\nu_{\sigma}(t+ dt) \rangle\ri|^{2} = 1-\frac{dt^{2}}{\hbar^{2}}
(\Omm )^{2} \sin^{2}\theta\,\cos^{2}\theta\,+\,O(dt^{3})\,, \qquad
\sigma = e, \mu\,,
\eea
where we have used the equations
\bea\label{:H:e}
\,\langle\nu_{e}(t)|\, H\,|\nu_{e}(t)\rangle &=& \omega_{1}\,
\cos^{2}\theta\, +\, \omega_{2}\,\sin^{2}\theta\,,
\\ [2mm]\label{:H:mu}
\,\langle\nu_{\mu}(t)|\,
H\,|\nu_{\mu}(t)\rangle &=&
\omega_{2}\,\cos^{2}\theta \,+\, \omega_{1}\, \sin^{2}\theta \,,
\eea
and
\bea
\,\langle\nu_{e}(t)|\,
 H ^2\,|\nu_{e}(t)\rangle &=&
\omega_{1}^{2}\,\cos^{2}\theta\,
+\, \omega_{2}^{2}\, \sin^{2}\theta\, ,
\\ [2mm]
\,\langle\nu_{\mu}(t)|\,
 H ^2\,|\nu_{\mu}(t)\rangle &=&\omega_{2}^{2}\, \cos^{2}\theta\,
+ \,\omega_{1}^{2}\,\sin^{2}\theta\, .
\eea
From Eqs.(\ref{:H:e-mu}) and (\ref{deltaQM}) it follows:
\bea \label{deltaEQM}
&& \Delta E_{e,e} = \Delta E_{\mu,\mu} =  \Delta E_{e,\mu} = \Delta E_{\mu,e}.
\eea
Eq.(\ref{deltaEQM}) implies that $\Delta E_{e,\mu} $ in Eq.(\ref{sQM})
(see also Eq.(\ref{:H:e-mu})) is nothing but the energy
uncertainty (variance) given in Eq.(\ref{deltaQM}).
We also have
\bea\label{modulusQMem} |\langle \nu_{e}(t)
|\nu_{\mu}(t+ dt) \rangle|^{2} = |\langle \nu_{\mu}(t) |\nu_{e}(t+
dt) \rangle|^{2} = \frac{dt^{2}}{\hbar} \Delta E_{e, \mu} ^{2}\, +
\, O(dt^{3})\,.
\eea
 The Fubini-Study metric
\cite{Anandan:1990fq} is defined as follows \bea\label{Fubini}
ds^{2}\,=\,2\,g_{\mu\bar{\nu}}\,dZ^{\mu}\,d\bar{Z}^{\nu}\,=
\,4\,(1\,-\,|\langle \nu_{\sigma}(t) |\nu_{\sigma}(t+ dt)
\rangle|^{2})\,, \eea where $Z^{\mu}$ are coordinates in the
projective Hilbert space $\cal{P}$, which is the set of rays of the
Hilbert space $\cal{H}$. From Eqs.(\ref{modulusQM}), (\ref{deltaQM})
and (\ref{Fubini}), we have the infinitesimal geodetic distance
between the points $\Pi(|\nu_{e}(t)\rangle)$ and
$\Pi(|\nu_{e}(t+dt)\rangle)$  in the space $\cal{P}$
\bea \label{lineFS}
ds\,=\,2\,\frac{\Delta
E_{\sigma,\sigma}\,dt}{\hbar}\,=\,2\,\frac{\Omm
\,\sin\theta\,\cos\theta\, \,dt}{\hbar}\,.
\eea
In the case of the
neutrino mixing, the above defined Fubini-Study metric is the usual
metric on a sphere of unitary radius:
$ds^{2}\,=\,d\Theta^{2}\,+\,\sin^{2}\Theta\,  d\varphi^{2}$, with
$\Theta =2\,\theta$ ($\theta=$ mixing angle) and $\Theta \in
[0,\pi]$.
Since $\theta$ is constant, we have $ds\,=\,\sin 2 \theta\,  d\varphi$
and, by comparison with Eq.(\ref{lineFS}),
 $ d\varphi\,=\,\frac{\Omm }{\hbar}\,dt\,$. We thus have
\bea\label{faseQM} s =
\int_{0}^{2n\pi}\sin 2\theta \, d\varphi\,=\,2 \,n \pi \sin 2\theta\,.
\eea
Eq.(\ref{faseQM}) coincides with Eq.(\ref{sQM}), which  thus
represents the distance between neutrino evolution states, as
measured by the Fubini-Study metric, in the projective Hilbert space
$\cal{P}$.

The case of three and four flavor mixing has been considered in
Refs.\cite{Wang:2000ep,Law:2007fb} where it has been shown that a generalization
\cite{Pancharatnam,Samuel,Polavieja} of the geometric phase
 to non-cyclic evolution (non-cyclic phase or Pancharatnam phase) needs to be used in order
to capture the geometric aspects of the neutrino phase in such cases.
The definition for the non-cyclic phase adopted in Ref.\cite{Wang:2000ep} is
\bea\label{noncycl}
\beta = Arg \lf( \langle \nu_{\sigma}(0)|\exp \lf[\frac{i}{\hbar}\int_{0}^{t} \langle E (t') \rangle d t'\ri]
|\nu_{\sigma}(t)\rangle\ri)\,
\eea
where, for example, in the case of a three-flavor electron neutrino state, $\sigma=e$,
\bea
|\nu_{e}(t)\rangle = e^{-i \omega_{1}t} \cos \theta_{12} \cos \theta_{13}\, |\nu_{1}\rangle
+ e^{-i \omega_{2}t} \sin \theta_{12} \cos \theta_{13}\, |\nu_{2}\rangle
+ e^{-i \omega_{3}t} e^{-i \delta} \sin \theta_{13}\, |\nu_{3}\rangle\,,
\eea
with $\theta_{12}$ and $\theta_{13}$ mixing angles;  $\delta$ is the
$CP$ violating phase and $\langle E \rangle (t)$ is given by
\bea
\langle E \rangle (t) =  \omega_{1} \cos^{2} \theta_{12} \cos^{2} \theta_{13}
+ \omega_{2} \sin^{2} \theta_{12} \cos^{2} \theta_{13} +  \omega_{2} \sin^{2} \theta_{13}\,,
 \eea
from which $\beta_{ee}$ is calculated \cite{Wang:2000ep}.

\section{Non-cyclic phases for neutrino oscillations in QFT}

We now study the Aharonov-Anandan geometric invariant in the context
of QFT. For simplicity, we study only the case of two flavor mixing;
three flavor mixing including CP violation will be analyzed
elsewhere.

In a standard notation, the Dirac neutrino fields $\nu_{1}(x)$ and
$\nu_{2}(x)$ with definite masses $m_{1}$ and $m_{2}$, respectively,
are written as
\bea\label{freefi}
 \nu _{i}(x)=\frac{1}{\sqrt{V}}{\sum_{{\bf k} ,
r}} \left[ u^{r}_{{\bf k},i}\, \al^{r}_{{\bf k},i}(t) + v^{r}_{-{\bf
k},i}\, \bt^{r\dag}_{-{\bf k},i}(t) \ri] e^{i {\bf k}\cdot{\bf
x}},\text{ \qquad \qquad }i=1,2, \eea
with $ \al_{{\bf k},i}^{r}(t)=\al_{{\bf k},i}^{r}\, e^{-i\omega
_{k,i}t}$, $ \bt_{{\bf k},i}^{r\dag}(t) = \bt_{{\bf k},i}^{r\dag}\,
e^{i\omega_{k,i}t},$ and $ \omega _{k,i}=\sqrt{{\bf k}^{2} +
m_{i}^{2}}.$ The operator $\alpha ^{r}_{{\bf k},i}$ and $ \beta ^{r
}_{{\bf k},i}$, $ i=1,2 \;, \;r=1,2$, are the annihilator operators
for the vacuum state $|0\rangle_{1,2}\equiv|0\rangle_{1}\otimes
|0\rangle_{2}$: $\alpha ^{r}_{{\bf k},i}|0\rangle_{12}= \beta ^{r
}_{{\bf k},i}|0\rangle_{12}=0$. The above fields and wavefunctions
satisfy standard anti-commutation, orthonormality and completeness
relations (see Ref.\cite{BV95}).

The field mixing relations are
 \bea\label{mix} \nu _{e}(x) &=&\cos \theta\,\nu_{1}(x) +\sin \theta\,\nu _{2}(x)
\\[2mm]
\nu _{\mu }(x)
&=&-\sin \theta\,\nu _{1}(x) +\cos \theta\,\nu _{2}(x) \eea
where $\nu_{e}(x)$ and $\nu_{\mu}(x)$ are the Dirac neutrino fields
with definite flavors. The generator of these mixing transformations
is given by \cite{BV95}
\bea\label{generator12}
 G(\te,t) = exp\left[\theta \int
d^{3}{\bf x} \left(\nu_{1}^{\dag}(x) \nu_{2}(x) - \nu_{2}^{\dag}(x)
\nu_{1}(x) \right)\right]\;, \eea
\bea \label{mixG} \nu_{e}(x)  &=& G^{-1}(\te,t)\; \nu_{1}(x)\;
G(\te,t)
\\[2mm]
\nu_{\mu}(x) &=& G^{-1}(\te,t)\; \nu_{2}(x)\; G(\te,t)~. \eea

At finite volume, $G(\te,t)$ is an unitary operator,
$G^{-1}(\te,t)=G (- \te,t)=G^{\dag}(\te,t)$, preserving the
canonical anticommutation relations. The generator $G^{-1}(\te,t)$
maps the Hilbert space ${\cal H}_{1,2}$ for $\nu_{1}, \nu_{2}$
fields  to the Hilbert spaces for flavor fields ${\cal H}_{e,\mu}$~:
$G^{-1}(\te,t): {\cal H}_{1,2} \mapsto {\cal H}_{e,\mu}.$ In
particular, for the vacuum $|0 \rangle_{1,2}$ we have, at finite
volume $V$:
\bea\label{flavvac}
 |0(t) \rangle_{e,\mu} = G^{-1}(\te,t)\;
|0 \rangle_{1,2}\;. \eea
$|0 \rangle_{e,\mu}(t)$ is the vacuum for ${\cal H}_{e,\mu}$, which we
will refer to as the flavor vacuum. It is annihilated by the
annihilation operators of $\nu_{e}(x)$ and $\nu_{\mu}(x)$ neutrinos,
$\alpha _{{\bf k},\sigma}^{r}(t)|0(t) \rangle_{e,\mu} = 0 = \beta
_{{\bf k},\sigma}^{r}(t)|0(t) \rangle_{e,\mu} $, with $(\sigma,i)=(e,1)
, (\mu,2)$ and
\begin{eqnarray}\label{flavannich}
\alpha _{{\bf k},\sigma}^{r}(t) &\equiv &G^{-1}(\te,t)\;\alpha
_{{\bf k},i}^{r}(t)\;G(\te,t),
\\[2mm]
\beta _{{\bf k},\sigma}^{r}(t) &\equiv &G^{-1}(\te,t)\;\beta _{{\bf
k},i}^{r}(t)\;G(\te,t).
\end{eqnarray}

The non-trivial structure of the flavor vacuum is such that even in the simplest
two flavor case, flavor neutrino states have a multiparticle component
which makes non-cyclic the time evolution associated to them.
Indeed at time t, the flavor states in the reference frame
for which ${\bf k}=(0,0,|{\bf k}|)$ are:
\bea \non\label{h1(t)} |\nu_{{\bf k},e}^{r}(t) \rangle & \equiv & \alpha_{{\bf
k},e}^{r \dag}(t)|0(t)\rangle_{e,\mu} = e^{-i\, :H:\, t} |\nu_{{\bf
k},e}^{r}(0) \rangle,
\\\non
& = &  e^{-i \omega_{k,1} t} \left[ \cos\theta\,\alpha_{{\bf
k},1}^{r \dag} + |U_{\bf k}|e^{-i \Ommk t}\;
\sin\theta\;\alpha_{{\bf k},2}^{r \dag} - \epsilon^r \; |V_{\bf
k}|e^{-i \Ompk t} \,\sin\theta \; \alpha_{{\bf k},1}^{r
\dag}\alpha_{{\bf k},2}^{r \dag} \beta_{-{\bf k},1}^{r \dag} \right]
\\
& \times & G_{{\bf k},s \neq r}^{-1}(\theta,t) \prod_{{\bf p}\neq{\bf
k}} G_{\bf p}^{-1}(\theta,t)|0\rangle_{1,2}\,,
\\ [2mm] \non\label{h2(t)}
|\nu_{{\bf k},\mu}^{r}(t) \rangle &\equiv& \alpha_{{\bf k},\mu}^{r
\dag}(t)|0(t)\rangle_{e,\mu} = e^{-i\, :H:\, t} |\nu_{{\bf k},\mu}^{r}(0) \rangle,
\\\non
& = & e^{-i \omega_{k,2} t} \left[ \cos\theta\,\alpha_{{\bf k},2}^{r
\dag} - |U_{\bf k}|e^{i  \Ommk t}\;\; \sin\theta\;\alpha_{{\bf
k},1}^{r \dag} + \epsilon^r \; |V_{\bf k}|e^{- i \Ompk t}
\,\sin\theta \; \alpha_{{\bf k},1}^{r \dag}\alpha_{{\bf k},2}^{r
\dag} \beta_{-{\bf k},2}^{r \dag} \right]
\\
& \times & G_{{\bf k},s \neq r}^{-1}(\theta,t) \prod_{{\bf p}\neq{\bf
k}} G_{\bf p}^{-1}(\theta,t)|0\rangle_{1,2}\,,
 \eea
where  $\Ompk\equiv \omega_{k,2} + \omega_{k,1} $, $\Ommk \equiv
\omega_{k,2} - \omega_{k,1}$, and
\bea \label{Hnorm}:H: \,=\, H - _{1,2} \langle 0|H|0\rangle_{1,2} =
H + \, 2\int d^{3}{\bf k} \, \Ompk = \sum_{i}\sum_{r}\int d^{3}{\bf
k}\,\omega_{k,i}[\alpha_{{\bf k},i}^{r \dag}\alpha_{{\bf k},i}^{r}+
\beta_{{\bf k},i}^{r \dag}\beta_{{\bf k},i}^{r}] ~, \eea is the
Hamiltonian normal ordered with respect to the vacuum
$|0\rangle_{1,2}\,.$ It satisfies Eqs.(\ref{Hmass-e}) -
(\ref{Hmass-multi2m}) given in Appendix B, and
  $ :H: |\nu_{i} \rangle = \omega_{k,i} |\nu_{i}
\rangle$, with $i=1,2$.
We have used the notation $G (\theta,t) =
\prod_{\bf p} G_{{\bf p}} (\theta,t) = \prod_{\bf p} \prod_{s}
G_{{\bf p},s} (\theta,t)$ (cf. Eq. (\ref{generator12})).
Note that in the flavor states, the multi-particle components
disappear in the relativistic limit $|{\bf k}|\gg \sqrt{m_1m_2}\,$,
where  $\left| U_{\mathbf{k}}\right| ^{2}\rightarrow 1$ and $\left|
V_{\mathbf{k}}\right| ^{2}\rightarrow 0$ and the quantum mechanical
Pontecorvo's states are recovered.

Eqs.(\ref{h1(t)}), (\ref{h2(t)}) show that the non-cyclic time
evolution of mixed neutrino states is due to the presence of two
oscillation frequencies, namely $\Omp$ and $\Omm$.
Note however that
%we cannot calculate the geometric phases by using the procedure
%presented in Ref.\cite{Blasone:1999tq} (see Section II). Indeed,
%such a procedure is only applicable for cyclic evolution. One could
%then think to proceed in analogy with the three flavor case as done
%in Ref.\cite{Wang:2000ep} (see Section II). However,
the definition of
the geometric phase given in Eq.(\ref{noncycl}) is not applicable in the
QFT mixing formalism, since quantities like
$\langle \nu_{\sigma}(t) |\nu_{\sigma}(t^{\prime}) \rangle$, with
$t\neq t^{\prime}$, are zero in the infinite volume limit \cite{{Blasone:2005ae}}.
On the other hand, the geometric invariant defined in Ref.\cite{Anandan:1990fq}
(see  Eq.(\ref{sQM})) is suitable for the present case since it is
well defined in the case of non-cyclic time
evolution and does not involve products of states at different times. We thus consider the quantities
\bea\label{s-st} s_{\sigma,\tau}(t) & = & 2 \int_{0}^{t} \Delta
E_{\sigma,\tau}\, dt ~, \eea
where  $\Delta E \equiv \Delta E_{{\bf k}}^{r}$ and $\sigma,\tau$ are labels specifying the states used in
computing the uncertainties $\Delta E_{\sigma,\tau}$ in the integrals.

We first
compute  $\Delta E_{\sigma,\sigma}$ with $ \sigma =
e, \mu\,$ by using $:H:$. We have
\bea \Delta E_{\sigma,\sigma}^{2} & = & \langle\nu_{{\bf k,\sigma}}^{r}(t)|\, (:
H :)^{2}\,|\nu_{{\bf k,\sigma}}^{r}(t)\rangle - \langle\nu_{{\bf k,\sigma}}^{r}(t)| : H
:|\nu_{{\bf k,\sigma}}^{r}(t)\rangle^{2}, \quad \quad \sigma = e, \mu\, ~.
\eea
By using Eqs.(\ref{Hmass-e}), (\ref{Hmass2-e}), and
(\ref{Hmass-mu}), (\ref{Hmass2-mu}),  we obtain
\bea \label{33} \Delta E_{e,e}^{2} & = &
\sin^{2}\theta\,\cos^{2}\theta\,\lf[(\Ommk)^{2} + 4\, \omega_{k,1}
\omega_{k,2} |V_{\bf k}|^{2}\ri] + 4 \,\omega^{2}_{k,1}\,
\sin^{4}\theta\,|U_{\bf k}|^{2}|V_{\bf k}|^{2} \,,
\\ [2mm]
\label{34} \Delta E_{\mu,\mu}^{2} & = &
\sin^{2}\theta\,\cos^{2}\theta\, \lf[(\Ommk)^{2} + 4\, \omega_{k,1}
\omega_{k,2} |V_{\bf k}|^{2}\ri] + 4\,
\omega^{2}_{k,2}\,\sin^{4}\theta\,|U_{\bf k}|^{2}|V_{\bf k}|^{2} \,.
\eea
In analogy with Eq.(\ref{:H:e-mu}) defined in QM, $\Delta E_{e,\mu}$
in QFT is given by \bea\label{DeltaEemu} \Delta E_{e,\mu}  =
\langle \nu_{{\bf k},e}^{r}(t)| :H:|\nu_{{\bf
k},\mu}^{r}(t)\rangle &=& \langle\nu_{{\bf k},\mu}^{r}(t)| : H
:|\nu_{{\bf k},e}^{r}(t)\rangle = \Ommk\,\sin \theta\, \cos
\theta\,|U_{\bf k}|\,. \eea

By defining, at time $t$, the multi-particle flavor states (their explicit expressions are given in Appendix A):
\bea
\label{multi1} |\nu_{{\bf k},e \bar{e} \mu}^{r}(t)\rangle & \equiv &
 \alpha_{{\bf k},e}^{r \dag}(t)\,
\beta_{-{\bf k},e}^{r \dag}(t)\, \alpha_{{\bf k},\mu}^{r
\dag}(t)\,|0(t)\rangle_{e,\mu} \,,
\\ [2mm]
\label{multi2}|\nu_{{\bf k},\mu  \bar{\mu} e}^{r}(t)\rangle & \equiv &
\alpha_{{\bf k},\mu}^{r \dag}(t)\, \beta_{-{\bf k},\mu}^{r \dag}(t)\,
\alpha_{{\bf k},e}^{r \dag}(t)\,|0(t)\rangle_{e,\mu}\,,
\eea
we have also the following non-zero expectation values:
\bea\label{deltaL1}  \Delta E_{\mu \bar{e} e ,e} & = &
 \langle\nu_{{\bf k},{\mu} \bar{e} e}^{r}(t)|
:H:|\nu_{{\bf k},e}^{r}(t)\rangle \, , \quad ~~ \Delta E_{e \bar{\mu} \mu ,e}
= \langle\nu_{{\bf k},e \bar{\mu} \mu}^{r}(t)|:H:|\nu_{{\bf k},e}^{r}(t)\rangle \,,
\\ [2mm]
\label{deltaL} \Delta E_{\mu \bar{e} e ,\mu} & = &
 \langle\nu_{{\bf k},{\mu} \bar{e} e}^{r}(t)|
:H:|\nu_{{\bf k},\mu}^{r}(t)\rangle \, , \quad ~~ \Delta E_{e \bar{\mu} \mu
,\mu}  =  \langle\nu_{{\bf k},{e \bar{\mu} \mu}}^{r}(t)|
:H:|\nu_{{\bf k},{\mu}}^{r}(t)\rangle \, , \eea
whose explicit expressions are given in Appendix B.

Let us now note that $\Delta E_{e,e}^{2}$ and  $\Delta E_{\mu,\mu}^{2}$  can be also obtained as follows
\bea\label{deltaE1}
\Delta E_{e,e}^{2} &=&\Delta E_{e , \mu}^{2}\,+\,
\Delta E_{\mu \bar{e} e ,e}^{2}\,+\,
 \Delta E_{e \bar{\mu} \mu ,e}^{2}\,,
 \\ [2mm]\label{deltaE2}
 \Delta E_{\mu,\mu}^{2} &=&\Delta E_{e , \mu}^{2}\,+\,
\Delta E_{\mu \bar{e} e ,\mu}^{2}\,+\,
 \Delta E_{e \bar{\mu} \mu ,\mu}^{2}\,.
\eea
Eqs.(\ref{deltaE1}), (\ref{deltaE2}) represent a generalization
of the relation (\ref{deltaEQM}) to the case of QFT flavor states
taking into account the multiparticle components due to the
condensate structure of the flavor vacuum.

The explicit expressions for $s_{\sigma,\tau}$, with $\sigma,\tau = e,\mu,e \bar{\mu} \mu,\mu \bar{e} e $
 are given by:
\bea\label{se} s_{e,e}(t) & = & 2\, t  \,\sin \theta\,\sqrt{
\cos^{2}\theta\,\lf[(\Ommk)^{2} + 4 \,\omega_{k,1} \omega_{k,2}
|V_{\bf k}|^{2}\ri] + 4\, \omega^{2}_{k,1}\,\sin^{2}\theta\,|U_{\bf
k}|^{2}|V_{\bf k}|^{2}} \,,
\\[2mm]\label{smu}
s_{\mu,\mu}(t)  & = & 2\, t  \,\sin \theta\,\sqrt{
\cos^{2}\theta\,\lf[(\Ommk)^{2} + 4\, \omega_{k,1} \omega_{k,2}
|V_{\bf k}|^{2}\ri] + 4 \,\omega^{2}_{k,2}\,\sin^{2}\theta\,|U_{\bf
k}|^{2}|V_{\bf k}|^{2}} \,, \eea \bea \label{semu} &&s_{e , \mu}(t)
=  \Ommk\,t\,\sin 2\theta\,|U_{\bf k}|\,, \qquad\;\qquad
\;\qquad\,\,\,\,\,\,\,\,\,\, s_{\mu \bar{e} e, e}(t)  \, = \, s_{e
\bar{\mu} \mu, \mu}(t)  \,=\, \epsilon^r \,\Ompk \,t\,\sin
2\theta\,|V_{\bf k}|\,,
\\ [2mm]\label{smumue}
&&s_{e \bar{\mu} \mu, e}(t)   =   4\, \epsilon^r \,\omega_{k,1}
\,t\,\sin^{2}\theta\,
|U_{\bf k}|\,|V_{\bf k}|\,,
\quad\;\quad \,\,\,\,\,\,
s_{\mu \bar{e} e, \mu}(t)  \, = \, - 4\, \epsilon^r \,\omega_{k,2}\,t\,\sin^{2}\theta\,
|U_{\bf k}|\,|V_{\bf k}|\,.
\eea

From Eqs.(\ref{se})-(\ref{smumue})  we see that  in the
relativistic limit, ${\bf k}\gg \sqrt{m_{1}m_{2}}$, where $|V_{\bf
k}|\rightarrow 0$, $|U_{\bf k}|\rightarrow 1$, we have  $s_{\mu
\bar{e} e, e} = s_{e \bar{\mu} \mu, e} = s_{\mu \bar{e} e, \mu} =
s_{e \bar{\mu} \mu, \mu} = 0$. In such a limit, from Appendix B and
Eqs. (\ref{33}), (\ref{34}), we have $\Delta E_{e,e} = \Delta
E_{\mu,\mu}= \Delta E_{e, \mu}= \Ommk \sin \theta\,\cos \theta\,. $
In particular, if the time $t$ is set $t = 2 n  \pi /\Ommk $, the
quantum mechanical result is consistently recovered and the
geometric invariants $s_{e,e} = s_{\mu,\mu} = s_{e, \mu} = 2 n \pi
\sin 2\theta\,$ coincide  with the one given in Eq.(\ref{sQM}).

We point out that, since $|0\rangle_{1,2}$ and
$|0\rangle_{e,\mu}$ are unitary inequivalent states in the infinite
volume limit, two different normal orderings must be defined,
respectively with respect to the vacuum $|0\rangle_{1,2}$ for fields
with definite masses, as usual denoted by $:...:$,  and with respect
to the vacuum for fields with definite flavor $|0\rangle_{e,\mu}$,
denoted by $::...::$~.
The uncertainties $\Delta E_{\sigma,\tau}$ can be then computed by using $:H:$ as done above
or with $\nof H \nof$. The Hamiltonian normal ordered with respect
to the vacuum $|0\rangle_{e,\mu}$ is given by
\bea\label{Hflav} \nof H \nof\, \equiv \,  H \, -\, {}_{e,\mu}\lan
0| H | 0 \ran_{e,\mu}\, = H \, +\, 2\int d^{3}{\bf k} \, \Ompk \,(1
- 2\,|V_{\bf k}|^{2} \sin^{2}\theta)\,. \eea
Considering now the expectation values of $\nof H
\nof$ on the flavor states given in
%satisfies Eqs.(\ref{Hflav-e}) - (\ref{Hflav-multi3m}) given in
Appendix B, we have
\bea \Delta E_{e,\mu} & = & \langle\nu_{{\bf k},e}^{r}(t)|
:H:|\nu_{{\bf k},\mu}^{r}(t)\rangle \, = \, \langle\nu_{{\bf k},e}^{r}(t)| \nof H
\nof|\nu_{{\bf k},\mu}^{r}(t)\rangle\,. \eea
On the other hand, defining the uncertainties $\Delta \widetilde{E}_{\sigma,\sigma}$
 as
\bea \Delta \widetilde{E}_{\sigma,\sigma}^{2} =
\langle\nu_{{\bf k},\sigma}^{r}(t) | (\nof H
\nof)^{2}|\nu_{{\bf k},\sigma}^{r}(t)\rangle - \langle\nu_{{\bf k},\sigma}^{r}(t)| \nof
H \nof|\nu_{{\bf k},\sigma}^{r}(t)\rangle^{2}, \qquad \sigma = e, \mu\,,
 \eea
and by using the relations in Appendix B, we have
$\Delta \widetilde{E}_{e,e}^{2}= \Delta E_{e,e}^{2}$ and $\Delta
\widetilde{E}_{\mu,\mu}^{2} =\Delta E_{\mu,\mu}^{2}$,
 that is, $\Delta E_{\sigma,\sigma}^{2}$
are independent of the normal ordering used, $:H:$ or $\nof H \nof$.
Moreover, by comparing the expectation values of $:H:$ and $\nof H
\nof$ presented in Appendix B, we obtain that $ \Delta E_{e,\mu}\,,
\Delta E_{\mu \bar{e} e ,e}\,, \Delta E_{e \bar{\mu} \mu ,e}\,,
\Delta E_{\mu \bar{e} e ,\mu}\,, \Delta E_{e \bar{\mu} \mu ,\mu}\,$
are also independent of the particular  normal ordering used.
This implies that the invariants of Eqs.(\ref{se})-(\ref{smumue}) are independent of
the normal ordering used.

\section{discussion and Conclusions }

Let us conclude the paper with some further comments.
It is interesting to define the operator $H'(t)$:
\bea \non\label{H'(t)}
H'(t) & \equiv & \sum_{r} \int d^{3}{\bf k}\,\Big[\omega_{e e}
 \lf(\alpha_{{\bf k},e}^{r \dag}(t)\alpha_{{\bf k},e}^{r}(t)\, +
 \, \beta_{-{\bf k},e}^{r \dag}(t)\beta_{-{\bf k},e}^{r}(t) \ri)
\,+\,\omega_{\mu \mu} \lf(\alpha_{{\bf k},\mu}^{r \dag}(t)
 \alpha_{{\bf k},\mu}^{r}(t) + \beta_{-{\bf k},\mu}^{r \dag}(t)\beta_{-{\bf k},\mu}^{r}(t) \ri)
 \\
& + & \omega_{\mu e}
\lf(\alpha_{{\bf k},e}^{r \dag}(t)\alpha_{{\bf k},\mu}^{r}(t)\, +
\, \alpha_{{\bf k},\mu}^{r \dag}(t) \alpha_{{\bf k},e}^{r}(t)\, +
 \, \beta_{-{\bf k},e}^{r \dag}(t)\beta_{-{\bf k},\mu}^{r }(t)\, +
\, \beta_{-{\bf k},\mu}^{r \dag}(t) \beta_{-{\bf k},e}^{r
}(t)\ri)\Big].
\eea
where
$\omega_{e e} \equiv \omega_{k,1}\,
\cos^{2}\theta \,+ \omega_{k,2}\, \sin^{2}\theta$, $\omega_{\mu \mu}
\equiv \omega_{k,1}\, \sin^{2}\theta \,+ \omega_{k,2}\,
\cos^{2}\theta$ , $ \omega_{\mu e} \equiv \Ommk\,\sin\theta
\cos\theta $.
We have
\bea\label{H'(t)e}
\langle\nu_{{\bf k},e}^{r}(t)|\,
H'(t)\,|\nu_{{\bf k},e}^{r}(t)\rangle & = &
\omega_{k,1} \cos^{2}\theta\, + \, \omega_{k,2}\sin^{2}\theta\,,
\\\label{H'(t)mu}
\langle\nu_{{\bf k},\mu}^{r}(t)|\,
H'(t)\,|\nu_{{\bf k},\mu}^{r}(t)\rangle &=& \omega_{k,1}\sin^{2}\theta\,+\,
\omega_{k,2} \cos^{2}\theta\,  ,
\\\label{H'(t)e-mu}
\langle\nu_{{\bf k},e}^{r}(t)|\, H'(t)\,|\nu_{{\bf
k},\mu}^{r}(t)\rangle &=& \Ommk\sin \theta \cos \theta\,, \eea
\bea \label{H'(t)multi1}\non
&&\langle\nu_{{\bf k},\mu \bar{e} e}^{r}(t)|\,
H'(t)\,|\nu_{{\bf k},e}^{r}(t)\rangle \, = \,
\langle\nu_{{\bf k},\mu \bar{e} e}^{r}(t)|\,
H'(t)\,|\nu_{{\bf k},\mu}^{r}(t)\rangle\,= \,\langle\nu_{{\bf k},e \bar{\mu} \mu}^{r}(t)|\,
H'(t)\,|\nu_{{\bf k},e}^{r}(t)\rangle \, = \,
\langle\nu_{{\bf k},e \bar{\mu} \mu}^{r}(t)|\,
H'(t)\,|\nu_{{\bf k},\mu}^{r}(t)\rangle = 0\,.
\\
\eea
From the above expectation values, we see that contributions from the flavor vacuum condensate
have been eliminated. Indeed,
Eqs.(\ref{H'(t)e})-(\ref{H'(t)e-mu}) coincide with Eqs.(\ref{:H:e}),
(\ref{:H:mu}) and (\ref{:H:e-mu}) derived in the QM case (see Section II).
Moreover the
uncertainties in the energy $H'(t)$ of the multi-particle states
(\ref{multi1}), (\ref{multi2}) are zero such as in QM.

An invariant analogous to the one introduced in Section II, can be
then defined as \bea\label{faseQFT} s'_{e} =s'_{\mu} =  2
\int_{0}^{n T} \Delta E'\, dt\, = 2 n \pi  \sin 2\theta\,, \eea
where $T = 2 n  \pi /\Ommk $ and \bea \non\label{deltaEP} \Delta
{E'}_{e,e}^{2} = \Delta {E'}_{\mu,\mu}^{2} = \Delta
{E'}_{e,\mu}^{2}&=& \langle\nu_{{\bf k},\sigma}^{r}(t)|\,
{H'}^{2}(t)\,|\nu_{{\bf k},\sigma}^{r}(t)\rangle - \langle\nu_{{\bf
k},\sigma}^{r}(t)|\,
 H'(t) \,|\nu_{{\bf k},\sigma}^{r}(t)\rangle^{2}
\\ & = &
\langle\nu_{{\bf k},e}^{r}(t)|\,  H'(t) \,|\nu_{{\bf
k},\mu}^{r}(t)\rangle^{2}\, = \,(\Ommk)^{2}
\sin^{2}\theta\,\cos^{2}\theta\,, \qquad \sigma = e, \mu\,.\eea

\medskip

In summary, in this paper we have calculated the non-cyclic phases
for neutrino oscillations in the context of QFT, for the case of two
flavors. In the relativistic limit, where the quantum mechanical
approximation holds, the QM geometric phase is recovered. The above
analysis is suitable for treatment of three flavor case (see
Ref.\cite{yBCV02}) where however, differences due to the presence of
CP violating phase are expected.

Questions not considered in the present paper, like the extension
of  the present and previous results to wave-packet formalism,
or the suggestion of experimental setups by means of which geometric phases
associated to neutrino oscillations could
be detected are certainly interesting and deserve a separate analysis.

\acknowledgments

We acknowledge partial financial support from MIUR and INFN.

\appendix

\section{Flavor fields and QFT flavor states}

By taking into account the relations Eqs.(\ref{freefi})-(\ref{flavannich}),
the flavor fields can be written as:
\begin{eqnarray}
\nu _{\sigma}({\bf x},t) &=&\frac{1}{\sqrt{V}}{\sum_{{\bf k},r} }
e^{i{\bf k.x}}\left[ u_{{\bf k},i}^{r} \alpha _{{\bf
k},\sigma}^{r}(t) + v_{-{\bf k},i}^{r} \beta _{-{\bf
k},\sigma}^{r\dagger }(t)\right] , \quad (\sigma,i)=(e,1) ~, (\mu,2)
\end{eqnarray}
In the reference frame such that ${\bf k}=(0,0,|{\bf k}|)$ the
annihilation operators of  $\nu_{e}(x)$ and $\nu_{\mu}(x)$ are
explicitly given by
%Due to the linearity of
%$G(\te,t)$, we can define the flavor annihilators, relative
%to the fields $\nu_{e}(x)$ and $\nu_{\mu}(x)$ at each time
%expressed as (we use $(\sigma,i)=(e,1) , (\mu,2)$):
%%
%\begin{eqnarray}\label{flavannich}
%\alpha _{{\bf k},\sigma}^{r}(t) &\equiv &G^{-1}_{\bf
%\te}(t)\;\alpha _{{\bf k},i}^{r}(t)\;G(\te,t),  \nonumber
%\\[2mm]
%\beta _{{\bf k},\sigma}^{r}(t) &\equiv &G^{-1}(\te,t)\;\beta
%_{{\bf k},i}^{r}(t)\;G(\te,t).
%\end{eqnarray}
%
\bea\label{annihilator}
\alpha^{r}_{{\bf
k},e}(t)&=&\cos\theta\;\alpha^{r}_{{\bf
k},1}(t)\;+\;\sin\theta\;\left( |U_{{\bf k}}|\; \alpha^{r}_{{\bf
k},2}(t)\;+\;\epsilon^{r}\; |V_{{\bf k}}|\; \beta^{r\dag}_{-{\bf
k},2}(t)\right)
\\
\alpha^{r}_{{\bf k},\mu}(t)&=&\cos\theta\;\alpha^{r}_{{\bf
k},2}(t)\;-\;\sin\theta\;\left( |U_{{\bf k}}|\; \alpha^{r}_{{\bf
k},1}(t)\;-\;\epsilon^{r}\; |V_{{\bf k}}|\; \beta^{r\dag}_{-{\bf
k},1}(t)\right)
\\
\beta^{r}_{-{\bf k},e}(t)&=&\cos\theta\;\beta^{r}_{-{\bf
k},1}(t)\;+\;\sin\theta\;\left( |U_{{\bf k}}|\; \beta^{r}_{-{\bf
k},2}(t)\;-\;\epsilon^{r}\; |V_{{\bf k}}|\; \alpha^{r\dag}_{{\bf
k},2}(t)\right)
\\
\beta^{r}_{-{\bf k},\mu}(t)&=&\cos\theta\;\beta^{r}_{-{\bf
k},2}(t)\;-\;\sin\theta\;\left( |U_{{\bf k}}|\; \beta^{r}_{-{\bf
k},1}(t)\;+\;\epsilon^{r}\; |V_{{\bf k}}|\; \alpha^{r\dag}_{{\bf
k},1}(t)\right), \eea
with $\epsilon^{r}=(-1)^{r}$ and
\bea\label{Vk2} \non |U_{{\bf k}}| \equiv  u^{r\dag}_{{\bf k},i}
u^{r}_{{\bf k},j} = v^{r\dag}_{-{\bf k},i} v^{r}_{-{\bf k},j} ~,
\quad \quad |V_{{\bf k}}| \equiv  \epsilon^{r}\; u^{r\dag}_{{\bf
k},1} v^{r}_{-{\bf k},2} = -\epsilon^{r}\; u^{r\dag}_{{\bf k},2}
v^{r}_{-{\bf k},1} \eea
where $i,j = 1,2$ and $ i \neq j$. We have:
\bea \label{Vk}
&& |U_{{\bf k}}|=\frac{|{\bf k}|^{2} +(\om_{k,1}+m_{1})(\om_{k,2}+m_{2})}{2
\sqrt{\om_{k,1}\om_{k,2}(\om_{k,1}+m_{1})(\om_{k,2}+m_{2})}} \quad  ;
\quad
|V_{{\bf k}}|=\frac{ (\om_{k,1}+m_{1}) - (\om_{k,2}+m_{2})}{2
\sqrt{\om_{k,1}\om_{k,2}(\om_{k,1}+m_{1})(\om_{k,2}+m_{2})}}\, |{\bf k}| \, ,
\eea
\bea|U_{{\bf k}}|^{2}+|V_{{\bf k}}|^{2}=1. \eea
The number of condensed neutrinos for each ${\bf k}$ is given by
\bea
 _{e,\mu}\langle 0| \al_{{\bf k},i}^{r \dag} \al^r_{{\bf k},i}
|0\rangle_{e,\mu}\,= \;_{e,\mu}\langle 0| \bt_{{\bf k},i}^{r \dag}
\bt^r_{{\bf k},i} |0\rangle_{e,\mu}\,=\, \sin^{2}\te\; |V_{{\bf
k}}|^{2} \;, \qquad i=1,2\,.
\eea
The explicit expression for $|0\rangle_{e,\mu}$  at time $t=0$ in
the reference frame for which ${\bf k}=(0,0,|{\bf k}|)$ is

\bea\non\label{0emu} |0\rangle_{e,\mu} &=& \prod_{r,{\bf k}}
\Big[(1-\sin^{2}\theta\;|V_{{\bf k}}|^{2})
-\epsilon^{r}\sin\theta\;\cos\theta\; |V_{{\bf k}}|
(\alpha^{r\dag}_{{\bf k},1}\beta^{r\dag}_{-{\bf k},2}+
\alpha^{r\dag}_{{\bf k},2}\beta^{r\dag}_{-{\bf k},1})+
\\
 &+&\epsilon^{r}\sin^{2}\theta \;|V_{{\bf k}}||U_{{\bf
k}}|(\alpha^{r\dag}_{{\bf k},1}\beta^{r\dag}_{-{\bf k},1}-
\alpha^{r\dag}_{{\bf k},2}\beta^{r\dag}_{-{\bf k},2})
+\sin^{2}\theta \; |V_{{\bf k}}|^{2}\alpha^{r\dag}_{{\bf
k},1}\beta^{r\dag}_{-{\bf k},2} \alpha^{r\dag}_{{\bf
k},2}\beta^{r\dag}_{-{\bf k},1} \Big]|0\rangle_{1,2}\,.
 \eea
 Eq.(\ref{0emu}) exhibits the condensate
structure of the flavor vacuum $|0\rangle_{e,\mu}$. The important
point is that $_{1,2}\langle 0 |0 (t)\rangle_{e,\mu} \rar 0$, for
any $t$, in the infinite volume limit \cite{BV95}. Thus, in such a
limit the Hilbert spaces ${\cal H}_{1,2}$ and ${\cal H}_{e,\mu}$
turn out to be unitarily inequivalent spaces.

The explicit form of the multi-particle states
defined in Eqs.(\ref{multi1}), (\ref{multi2}) is:

 \bea\non
 |\nu_{{\bf k},e \bar{e} \mu}^{r}(t)\rangle &= &
 - \Big[ \cos\theta \;
\alpha_{{\bf k},1}^{r \dag} \alpha_{{\bf k},2}^{r \dag}
\beta_{-{\bf k},1}^{r \dag}\,e^{-i (2\omega_{k,1} + \omega_{k,2}) t}\,+\,
\epsilon^r \;|V_{\bf
k}|\;\sin\theta\;  \alpha_{{\bf k},1}^{r \dag}\,e^{-i \omega_{k,1} t}
\\ [2mm]&& +
 |U_{\bf k}|\;\sin\theta\; \alpha_{{\bf k},1}^{r \dag} \alpha_{{\bf
k},2}^{r \dag} \beta_{-{\bf k},2}^{r \dag}\, e^{-i (\omega_{k,1} + 2\omega_{k,2}) t} \Big]
 G_{{\bf k},s \neq r}^{-1}(\theta,t) \prod_{{\bf p}\neq{\bf k}} G_{\bf p}^{-1}(\theta,t)|0\rangle_{1,2}\,,
\\\non
|\nu_{{\bf k},\mu  \bar{\mu} e}^{r}(t)\rangle & = &
\Big[ \cos\theta \; \alpha_{{\bf k},1}^{r \dag} \alpha_{{\bf k},2}^{r \dag}
\beta_{-{\bf k},2}^{r \dag}\, e^{-i (\omega_{k,1} + 2\omega_{k,2}) t}\, - \,
\epsilon^r \;|V_{\bf
k}|\;\sin\theta\;  \alpha_{{\bf k},2}^{r \dag}\, e^{-i \omega_{k,2} t}
\\ [2mm]
& &-
 |U_{\bf k}|\;\sin\theta\; \alpha_{{\bf k},1}^{r \dag} \alpha_{{\bf
k},2}^{r \dag} \beta_{-{\bf k},1}^{r \dag}\, e^{-i (2\omega_{k,1} + \omega_{k,2}) t} \Big]
G_{{\bf k},s \neq r}^{-1}(\theta,t) \prod_{{\bf p}\neq{\bf k}} G_{\bf p}^{-1}(\theta,t)|0\rangle_{1,2}\,,
\eea

\section{Expectation values of $: H :$ and $\nof H \nof$}

The flavor states introduced in the Appendix A are used in
computing the following expectation values for the Hamiltonian $: H
:$~, $\nof H \nof$. We have:
\bea
\label{Hmass-e}
\langle\nu_{{\bf k},e}^{r}(t)|
: H :|\nu_{{\bf k},e}^{r}(t)\rangle &=&
\omega_{k,1}\, (\cos^{2}\theta\, +\, 2\, \sin^{2}\theta \,|V_{\bf k}|^{2})\,+\,
\omega_{k,2}\,\sin^{2}\theta\,,
\\
\label{Hmass-mu} \langle\nu_{{\bf k},\mu}^{r}(t)|: H :|\nu_{{\bf k},\mu}^{r}(t)\rangle
&=&  \omega_{k,2}\,
(\cos^{2}\theta\, +\, 2\, \sin^{2}\theta \,|V_{\bf k}|^{2}) \,+\,\omega_{k,1}\,\sin^{2}\theta\, ,
\eea
\bea \label{Hmass2-e} \langle\nu_{{\bf k},e}^{r}(t)|\, (: H
:)^2\,|\nu_{{\bf k},e}^{r}(t)\rangle &=& \omega_{k,1}^{2}\, (\cos^{2}\theta\,
+\, 4\,\sin^{2}\theta\, |V_{\bf k}|^{2})\, +\, \omega_{k,2}^{2}\,
\sin^{2}\theta\, +
\,4\,\omega_{k,1}\,\omega_{k,2}\sin^{2}\theta\,|V_{\bf k}|^{2}\,,
\\
\label{Hmass2-mu} \langle\nu_{{\bf k},\mu}^{r}(t)| (: H
:)^2|\nu_{{\bf k},\mu}^{r}(t)\rangle &=&
 \omega_{k,2}^{2}\,
 (\cos^{2}\theta\, +\, 4\,\sin^{2}\theta\, |V_{\bf k}|^{2})\,
+\,\omega_{k,1}^{2}\, \sin^{2}\theta\, +
\,4\,\omega_{k,1}\,\omega_{k,2}\sin^{2}\theta\,|V_{\bf k}|^{2}\,.
\eea
\bea
\label{Hmass-multi2}
\langle\nu_{{\bf k},e \bar{\mu} \mu}^{r}(t)|: H :|\nu_{{\bf k},e}^{r}(t)\rangle & = &
2\, \epsilon^r \,\omega_{k,1}\,\sin^{2}\theta\,|U_{\bf k}|\,|V_{\bf
k}|\,,
\\
\label{Hmass-multi1m}
\langle\nu_{{\bf k},\mu \bar{e} e}^{r}(t)|: H :|\nu_{{\bf k},\mu}^{r}(t)\rangle & = & -2\, \epsilon^r
\,\omega_{k,2}\,\sin^{2}\theta\,|U_{\bf k}|\,|V_{\bf k}|\, ,
\\
\label{Hmass-multi2m} \langle\nu_{{\bf k},e \bar{\mu} \mu}^{r}(t)|: H :|\nu_{{\bf k},\mu}^{r}(t)\rangle
& = & \langle\nu_{{\bf
k},\mu \bar{e} e}^{r}(t)|: H :|\nu_{{\bf
k},e}^{r}(t)\rangle\,=\,\epsilon^r \,\Ompk\,\sin \theta \,\cos
\theta\,|V_{\bf k}|\,,
 \eea

The Hamiltonian normal ordered with respect to the flavor vacuum
$\nof H \nof$  satisfies the following relations:
\bea \label{Hflav-e} \langle\nu_{{\bf k},e}^{r}(t)|\nof H
\nof|\nu_{{\bf k},e}^{r}(t)\rangle &=& \omega_{k,1}\, \cos^{2}\theta\, +
\,\omega_{k,2}\, \sin^{2}\theta\, (1 - 2\, |V_{\bf k}|^{2})\,,
\\
\label{Hflav-mu}
\langle\nu_{{\bf k},\mu}^{r}(t)|
\nof H \nof|\nu_{{\bf k},\mu}^{r}(t)\rangle &=&\omega_{k,2}\, \cos^{2}\theta\,+\,
\omega_{k,1}\, \sin^{2}\theta\, (1 - 2\, |V_{\bf k}|^{2})
\, ,
\\
\label{Hflav-emu} \langle\nu_{{\bf k},e}^{r}(t)| \nof H
\nof|\nu_{{\bf k},\mu}^{r}(t)\rangle  &=& \langle\nu_{{\bf
k},\mu}^{r}(t)| \nof H \nof|\nu_{{\bf k},e}^{r}(t)\rangle=
\Ommk\,\sin \theta\, \cos \theta \,|U_{\bf k}|\, , \eea
\bea
\label{Hflav2-e}
\langle\nu_{{\bf k},e}^{r}(t)|\,
(\nof H \nof)^2\,|\nu_{{\bf k},e}^{r}(t)\rangle & = &
\omega^{2}_{k,1}\,(\cos^{2}\theta\, + 4 \,\sin^{4}\theta\,|U_{\bf k}|^{2}\,|V_{\bf k}|^{2})
\, +\, \omega^{2}_{k,2}\,\sin^{2}\theta\,(1 - 4\,\sin^{2}\theta\,|U_{\bf k}|^{2}\,|V_{\bf k}|^{2})\, ,
\\
\label{Hflav2-mu}
\langle\nu_{{\bf k},\mu}^{r}(t)|\,
(\nof H \nof)^2\,|\nu_{{\bf k},\mu}^{r}(t)\rangle & = &
\omega^{2}_{k,2}\,(\cos^{2}\theta\, + 4 \,\sin^{4}\theta\,|U_{\bf k}|^{2}\,|V_{\bf k}|^{2})\,+\,
\omega^{2}_{k,1}\,\sin^{2}\theta\,(1 - 4\,\sin^{2}\theta\,|U_{\bf k}|^{2}\,|V_{\bf k}|^{2})
\,.
\eea
Finally we have:
\bea
&&\non \langle\nu_{{\bf k},e \bar{\mu} \mu}^{r}(t)|\nof H \nof|\nu_{{\bf k},e}^{r}(t)\rangle =
\langle\nu_{{\bf k},e \bar{\mu} \mu}^{r}(t)|: H :|\nu_{{\bf k},e}^{r}(t)\rangle \quad ; \quad
\langle\nu_{{\bf k},\mu \bar{e} e}^{r}(t)| \nof H \nof|\nu_{{\bf k},\mu}^{r}(t)\rangle =
\langle\nu_{{\bf k},\mu \bar{e} e}^{r}(t)|: H :|\nu_{{\bf k},\mu}^{r}(t)\rangle
\\
\\
&& \non\langle\nu_{{\bf k},\mu \bar{e} e}^{r}(t)|\nof H \nof|\nu_{{\bf k},e}^{r}(t)\rangle =
\langle\nu_{{\bf k},e \bar{\mu} \mu}^{r}(t)|\nof H \nof|\nu_{{\bf k},\mu}^{r}(t)\rangle =
\langle\nu_{{\bf k},e \bar{\mu} \mu}^{r}(t)|: H :|\nu_{{\bf k},\mu}^{r}(t)\rangle=
\langle\nu_{{\bf k},e \bar{\mu} \mu}^{r}(t)|: H :|\nu_{{\bf k},\mu}^{r}(t)\rangle
\\
\eea.


\begin{thebibliography}{99}

%\cite{Bilenky:1978nj}
\bibitem{Bilenky:1978nj}
  S.~M.~Bilenky and B.~Pontecorvo,
  % ``Lepton Mixing And Neutrino Oscillations,''
  Phys.\ Rept.\  {\bf 41}, 225 (1978).
  %%CITATION = PRPLC,41,225;%%

\bibitem{Bilenky}
S.~M.~Bilenky and S.~T.~Petcov,
%``Massive Neutrinos And Neutrino Oscillations,''
Rev.\ Mod.\ Phys.\ {\bf 59}, 671 (1987);
%%CITATION = RMPHA,59,671;%%
%\\ Super-Kamiokande Collaboration (Y. Fukuda et al.) {\it Phys. Rev.
%Lett.} {\bf 81}, 1562 (1998).
%\\ Kamiokande Collaboration (S.
%Hatakeyama et al.), {\it Phys. Rev. Lett.} {\bf 81}, 2016 (1998).
%
%
%%\cite{Gribov:1968kq}
%\bibitem{Gribov:1968kq}
%  V.~N.~Gribov and B.~Pontecorvo,
% %``Neutrino Astronomy And Lepton Charge,''
% Phys.\ Lett.\ B {\bf 28}, 493 (1969).
%  %%CITATION = PHLTA,B28,493;%%

  %\cite{Mohapatra:1991ng}
\bibitem{Mohapatra:1991ng}
  R.~N.~Mohapatra and P.~B.~Pal,
  %``Massive neutrinos in physics and astrophysics,''
  World Sci.\ Lect.\ Notes Phys.\  {\bf 41}, 1 (1991);
  %%CITATION = 00327,41,1;%%
 J.~N.~Bahcall,
{\it "Neutrino Astrophysics",} Cambridge Univ. Press, Cambridge, UK,
(1989).

\bibitem{BV95}
M.~Blasone and G.~Vitiello,
%``Quantum field theory of fermion mixing,''
Annals Phys.\ {\bf 244}, 283 (1995);
% [hep-ph/9501263].

%\cite{Blasone:1998hf}
\bibitem{BHV98}
  M.~Blasone, P.~A.~Henning and G.~Vitiello,
  %``The exact formula for neutrino oscillations,''
  Phys.\ Lett.\ B {\bf 451} 140 (1999).
  %[arXiv:hep-th/9803157].
  %%CITATION = HEP-TH 9803157;%%

\bibitem{BPT02}
M.~Blasone, P.~P.~Pacheco and H.~W.~Tseung,
%``Neutrino oscillations from relativistic flavor currents,''
Phys.\ Rev.\ D {\bf 67} (2003) 073011.
%%CITATION = HEP-PH 0212402;%%

 %\cite{Capolupo:2006et}
\bibitem{Capolupo:2006et}
A.~Capolupo, S.~Capozziello and G.~Vitiello,
  %``Dark energy explained by the mixing of neutrinos,''
 Phys.\ Lett.  A {\bf 363}, 53 (2007);
  %%CITATION = ASTRO-PH 0602467;%%
% \\
%\cite{Capolupo:2008rz}
%\bibitem{Capolupo:2008rz}
%  A.~Capolupo, S.~Capozziello and G.~Vitiello,
  %``Dark energy and particle mixing,''
  Phys.\ Lett. A  {\bf 373}, 601 (2009);
  %%CITATION = ARXIV:0809.0085;%%
% \\
%\cite{Capolupo:2007hy}
%\bibitem{Capolupo:2007hy}
 % A.~Capolupo, S.~Capozziello and G.~Vitiello,
  %``Dark energy, cosmological constant and neutrino mixing,''
 Int.\ J.\ Mod.\ Phys. A, {\bf 23}, 4979 (2008);
  %%CITATION = ARXIV:0705.0319;%%
% \\
%\cite{Capolupo:2006re}
%%\bibitem{Capolupo:2006re}
% % A.~Capolupo, S.~Capozziello and G.~Vitiello,
%  %``Dark energy induced by neutrino mixing,''
%  J.\ Phys.\ Conf.\ Ser.\  {\bf 67}, 012032 (2007);
% %[arXiv:hep-th/0612035].
%  %%CITATION = 00462,67,012032;%%
%%\\
%%\cite{Blasone:2007jm}
%%\bibitem{Blasone:2007jm}
%  M.~Blasone, A.~Capolupo, S.~Capozziello and G.~Vitiello,
%  %``Neutrino mixing, flavor states and dark energy,''
%  Nucl.\ Instrum.\ Meth.\  A {\bf 588}, 272 (2008).
% % [arXiv:0711.0939 [hep-th]].
%  %%CITATION = NUIMA,A588,272;%%
% %\\
 %\bibitem{Blasone:2004yh}
  M.~Blasone, A.~Capolupo, S.~Capozziello, S.~Carloni  and
G.~Vitiello
%``Neutrino mixing contribution to the cosmological constant,''
 Phys.\ Lett.\  A {\bf 323}, 182 (2004).
%%CITATION = GR-QC 0402013;%%


%\cite{Berry:1984jv}
\bibitem{Berry:1984jv}
  M.~V.~Berry,
  %``Quantal phase factors accompanying adiabatic changes,''
  Proc.\ Roy.\ Soc.\ Lond.\  A {\bf 392}, 45 (1984).
  %%CITATION = PRSLA,A392,45;%%

%\cite{Aharonov:1987gg}
\bibitem{Aharonov:1987gg}
  Y.~Aharonov and J.~Anandan,
  %``Phase Change During A Cyclic Quantum Evolution,''
  Phys.\ Rev.\ Lett.\  {\bf 58}, 1593 (1987).
  %%CITATION = PRLTA,58,1593;%%

%\cite{Anandan:1990fq}
\bibitem{Anandan:1990fq}
  J.~Anandan and Y.~Aharonov,
  %``GEOMETRY OF QUANTUM EVOLUTION,''
  Phys.\ Rev.\ Lett.\  {\bf 65}, 1697 (1990).
  %%CITATION = PRLTA,65,1697;%%




%\cite{Blasone:1999tq}
\bibitem{Blasone:1999tq}
  M.~Blasone, P.~A.~Henning and G.~Vitiello,
  %``Berry phase for oscillating neutrinos,''
  Phys.\ Lett.\  B {\bf 466}, 262 (1999).
  %[arXiv:hep-th/9902124].
  %%CITATION = PHLTA,B466,262;%%


%\cite{Capozziello:2000ga}
\bibitem{gaetano}
  S.~Capozziello and G.~Lambiase,
  %``Inertial effects on Berry's phase of neutrino oscillations,''
  Eur.\ Phys.\ J.\  C {\bf 16}, 155 (2000);
  %%CITATION = EPHJA,C16,155;%%
%\cite{Capozziello:2000ue}
  %``Berry's phase of neutrino oscillations in the presence of torsion,''
  Europhys.\ Lett.\  {\bf 52}, 15 (2000).
  %%CITATION = EULEE,52,15;%%


%\cite{He:2004zc}
\bibitem{He:2004zc}
  X.~G.~He, X.~Q.~Li, B.~H.~J.~McKellar and Y.~Zhang,
  %``Berry phase in neutrino oscillations,''
  Phys.\ Rev.\  D {\bf 72}, 053012 (2005).
  %%CITATION = PHRVA,D72,053012;%%


\bibitem{Pancharatnam}
S.~Pancharatnam, in Geometric Phases in Physics,
A.~D.~Shapere and F.~Wilczek Eds., World Scientific, Singapore (1989).


\bibitem{Samuel}
J.~Samuel and R.~Bhandari,
 Phys.\ Rev.\ Lett.\   {\bf 60}, 2339 (1988) .


\bibitem{Polavieja}
G.~G.~de Polavieja,
 Phys.\ Rev.\ Lett.\   {\bf 81}, 1 (1998) .


%\cite{Wang:2000ep}
\bibitem{Wang:2000ep}
  X.~B.~Wang, L.~C.~Kwek, Y.~Liu and C.~H.~Oh,
  %``Noncyclic geometric phase for neutrino oscillation,''
  Phys.\ Rev.\  D {\bf 63},  053003 (2001).
  %[arXiv:hep-ph/0006204].
  %%CITATION = PHRVA,D63,053003;%%

%\cite{Law:2007fb}
\bibitem{Law:2007fb}
  Z.~Y.~Law, A.~H.~Chan and C.~H.~Oh,
  %``Non-cyclic phase for 4-flavor neutrino oscillation,''
  Phys.\ Lett.\  B {\bf 648}, 289 (2007).
  %%CITATION = PHLTA,B648,289;%%


%\cite{Fujii:1999xa}
\bibitem{Fujii:1999xa}
K.~Fujii, C.~Habe and T.~Yabuki, Phys.\ Rev.\ D {\bf 59}, 113003
(1999); Phys.\ Rev.\ D {\bf 64}, 013011 (2001).

\bibitem{JM01}
C.R. Ji, Y. Mishchenko,
Phys. Rev. D {\bf 65}, 096015 (2002).




\bibitem{yBCV02}
M.~Blasone, A.~Capolupo and G.~Vitiello,
%``Quantum field theory of
%three flavor neutrino mixing and oscillations with CP violation,''
Phys.\ Rev.\ D {\bf 66}, 025033 (2002);
%%CITATION = HEP-TH 0204184;%%
%

%\cite{Blasone:2005ae}
\bibitem{Blasone:2005ae}
M.~Blasone, A.~Capolupo, F.~Terranova and G.~Vitiello,
%``Lepton charge and neutrino mixing in decay processes,''
Phys.\ Rev.\ D {\bf 72}, 013003 (2005).
  %[arXiv:hep-ph/0505178].
  %%CITATION = HEP-PH 0505178;%%

\bibitem{Blaspalm}
M.~Blasone and J.~Palmer,
%``Mixing and oscillations of neutral particles in quantum field theory,''
Phys.\ Rev.\ D {\bf 69}, 057301 (2004).
%[arXiv:hep-ph/0305257].
%%CITATION = HEP-PH 0305257;%%


\bibitem{BCRV01}
M.~Blasone, A.~Capolupo, O.~Romei and G.~Vitiello,
%``Quantum field theory of boson mixing,''
Phys.\ Rev.\ D {\bf 63}, 125015 (2001);
%[hep-ph/0102048].
%%CITATION = HEP-PH 0102048;%
%
%\bibitem{Capolupo:2004pt}
A.~Capolupo, C.~R.~Ji, Y.~Mishchenko and G.~Vitiello,
%``Phenomenology of flavor oscillations with non-perturbative effects from
%quantum field theory,''
Phys.\ Lett.\ B {\bf 594}, 135 (2004).
%[arXiv:hep-ph/0407166].
%%CITATION = HEP-PH 0407166;%%


\end{thebibliography}
\end{document}